# What Omicron Does, and How It Does It

J. C. Phillips

Dept. of Physics and Astronomy

Rutgers University, Piscataway, N. J., 08854

**Abstract**

Improvement of protein function by evolution (natural selection) is expected on general grounds, but even with the modern database positive proof has remained a difficult problem for theory. Here we extend our recent analysis of the evolution of CoV-1 to much more contagious CoV-2, to Omicron, which appears to be qualitatively different from other recent strains like Delta. Overall the synchronized dynamics of Omicron is more elaborate than CoV-2 or its variants like Delta. The surprising result is that while Omicron could be more contagious than even Delta, it is probably much less dangerous.

Introduction

Here we continue our analysis of Coronavirus spike dynamics using hydropathic profiling. Structural studies of the evolution of Coronavirus from CoV-1 to CoV-2 and its variants up to Delta have achieved many successes [1,2], but one of the most important features, the huge increase in contagiousness from CoV-1 to CoV-2, as continued by the Delta variant, has been puzzling. In earlier work, we were able to explain these increases, with only brief introductions in papers presenting results [3]. The new methods are difficult because they assemble simple methods from many special fields. The methods are based on the phase transition model, which is discussed in general terms in a widely read and cited physics article [4].

The surprising feature of the Omicron variant is that it involves ~ 30 mutations of CoV-2 [5]. CoV-2 itself involves ~ 300 mutations of the ~ 1200 amino acids in CoV-1, while the currently dominant Delta and similar variants involve ~ 10 mutations of CoV-2. Moreover, most of Omicron's 30 mutations are concentrated in the region associated with receptor attachment [5]; the mutations begin at 67 and end at 941. This suggests that the evolutionary mechanism that also explained details of the increase in contagiousness of the Delta and earlier variants of CoV-2 [6] has been replaced by a new mechanism.

Results

Earlier work showed that the increases in contagiousness that began with CoV-2 are correlated with the leveling of hydrophilic edges in hydropathic profiles Ψ(aa,W) based on the MZ hydrophobicity scale associated with second-order phase transitions [3,4,6]. . Level sets have been used by James Sethian, an applied mathematician, to study interface propagation and assembly in many systems, including fluid mechanics, semiconductor manufacturing, industrial inkjets and jetting devices, shape recovery in medicine, and medical and biomedical imaging [7]. He also uses the ancient Voronoi topological construction to separate and organize interfaces [8], much as [9] did to separate amino acid hydropathicities. These multiple and diverse applications suggest that level sets can also be useful in quantifying evolutionary trends in protein dynamics.

The early CoV work is summarized in Fig. 1, which shows the hydrophilic edge leveling as a function of W for CoV-1 to Delta. The leveling accelerates attachment through dynamical synchronization [3]. Maximum leveling occurs in CoV-2 and variants at W = 39. The important edges in Ψ(aa,39) are found in CoV-2 (Uniprot P0DTC2) at 7, 377; 1, 455; 8, 507; 2, 572; 3, 692; 4, 792; 5, 937; 6, 1156.

As before, we begin our search by profiling the hydropathic shape Ψ(aa,W) of the new Omicron strain. The best profiles were again obtained using the MZ scale [9] and a sliding window of width W = 39 [3,4,6]. Because Omicron originated in Botswana, which is adjacent to South Africa, we used as a reference sequence the South African strain B.1.351 [10]. The results in the attachment region are shown in Fig 2. There is a large qualitative difference between the two profiles. Fortunately the new features of the Omicron profile are explained by a small extension of the "level set" method [7,8] used to quantify the contagiousness of CoV-2 up through Delta including the South African strain B.1.351 [3,4,6]. The additional mutations in Omicron have leveled the two hydrophobic maxima (edges 7,8) near the static receptor binding domain [9]. The leveling of both groups is maximized at W = 39, as shown in Fig. 4. At W = 39 it is accurate to ~ 1%.

A curious feature of the Omicron data is that leveling occurs at W = 39 again, just as in CoV-2 and its variants like Delta. This suggests that there could be further leveling in Omicron, perhaps at a harmonic like W = 79. The profiles of Ψ(aa,79) for the South African strain B.1.351 and Omicron are compared in Fig. 5, which shows that there are four hydrophobic edges that are level in Omicron, but not in SA. Edges 9 (near 361) and 10 (near 629) are in S-1, whereas 11 (near 878) and 12 (near 1096) are in S-2. Edges 9-12 are level to 2%, which is excellent for such a large W. Values of W near 79 gave edges that are less level.

Discussion

CoV is cleaved into S-1 and S-2 near hydrophilic edge 3. CoV-2 is distinguished from CoV-1 by its greatly increased contagiousness, which was explained in [3] as a result of improved synchronization of six edges, including edges 1-3 in S-1, and edges 4-6 in S-2. As shown in Fig. 3, Omicron has lost the 4,5 edge synchronization, while improving the leveling of edges 1,2 and especially 3. It has also added a pair of level hydrophobic edges 7 and 8. Structural studies [11] have suggested that the narrow receptor binding domain, RBD, is supported by a core from 333-527, with a secondary structure of a twisted five-stranded antiparallel β sheet (β1, β2, β3, β4 and β7) with short connecting helices and loops. Between the β4 and β7 strands in the core, there is an extended insertion containing the short β5 and β6 strands, α4 and α5 helices and loops. This extended insertion is the RBM, which contains most of the contacting residues of SARS-CoV-2 that bind to ACE2. It appears that the hydrophobic edges 7 and 8 cut off the dynamics of this secondary core from the rest of S-1 below 333.

These 1-8 leveling differences in Ψ(aa,39) between CoV-2 South Africa and Omicron are numerous and accurate to ~ 1%. They are not only consistent with the observation that most of the Omicron mutations are concentrated in the receptor and receptor core region, but also the Omicron mutations in S-2 have desynchronized its dynamics. Moreover, as shown in Fig. 4, Omicron leveling, although removed from S-2 and concentrated in the core region, still uses the same nonlinear standing wave length W= 39 used by CoV-2 and its earlier variants, up to and including Delta. The same wave length synchronizes Omicron dynamics from 300 to 1200 aa. This is reminiscent of the dynamical water film waves that transport energy over a length of 1500 aa in the motor protein dynein [12]. In both cases the wave enables domains to move coherently across most of a topologically long, slender protein. Dynein moves along microtubules. Identifying these waves is possible because [9] discovered 20 fractals, one for each protein amino acid. Fractals are the latest feature of modern statistical physics. [13].

The unexpected leveling of edges 9-12 in Ψ(aa,79) can be interpreted as providing additional stability of S-1 (edges 9,10) against S-2 (edges 11,12) after the concentration in Ψ(aa,39) of level edges in S-1. Overall the synchronized dynamics of Omicron, which uses 12 level edges, is more elaborate than CoV-2 or its variants like Delta.

Evolution of CoV-2 occurred in a large, crowded Chinese city, whereas Delta probably evolved in a similarly large and crowded Indian city. There are no similar cities in Botswana, so how could such a remarkable strain as Omicron have evolved there? It might have happened during a heavy downpour, when natives crowded together for shelter [14]. Omicron could then have preferentially attached itself to the upper respiratory tracts, expelled to aerosols, and been rapidly spread through the country. Also the mutation D614G, previously proposed [3] as stabilizing CoV-2 in aerosols, is among the Omicron mutations. It has also been suggested that the insertion mutation 214EPE may have been obtained from the common cold [15], consistent with the downpour mechanism.

It is often supposed that a new strain must be more dangerous than an older strain, because it is more contagious. Omicron could be more contagious than Delta, although the latter has evolved in crowds in large cities. However, because Omicron is less synchronized in S-2, above the cleavage site, it should be less effective in forming trimeric fusion units [1], and therefore it could be less dangerous. Note: phase transition methods are useful in discussing CoV vaccines [16]. Specifically they explain the destructive interference of interdomain disulfide bonds when introduced into CoV to stabilize vaccines [2].

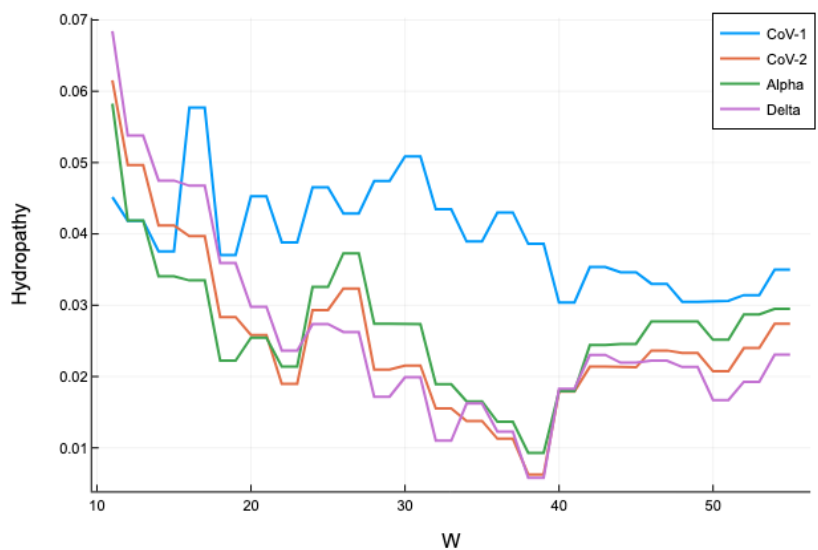

Fig. 1. The average deviations of six hydrophilic edges near 456, 572, 692, 792, 937 and 1163 in Ψ(aa,W) (or hydropathy) as a function of W. The 300 mutations in CoV-2 changed the leveling of these edges qualitatively, creating a single deep minimum at W = 39. The method used to estimate the optimal value of W used in [3] is replaced here by a much more thorough search for extrema using a Julia code that is available at https://github.com/ccc1685/SARS-CoV-2-Spike. Here Alpha refers to the UK strain, and Delta to the India strain. The smallest value of hydropathy in these two strains is about 10x smaller than in CoV-1.

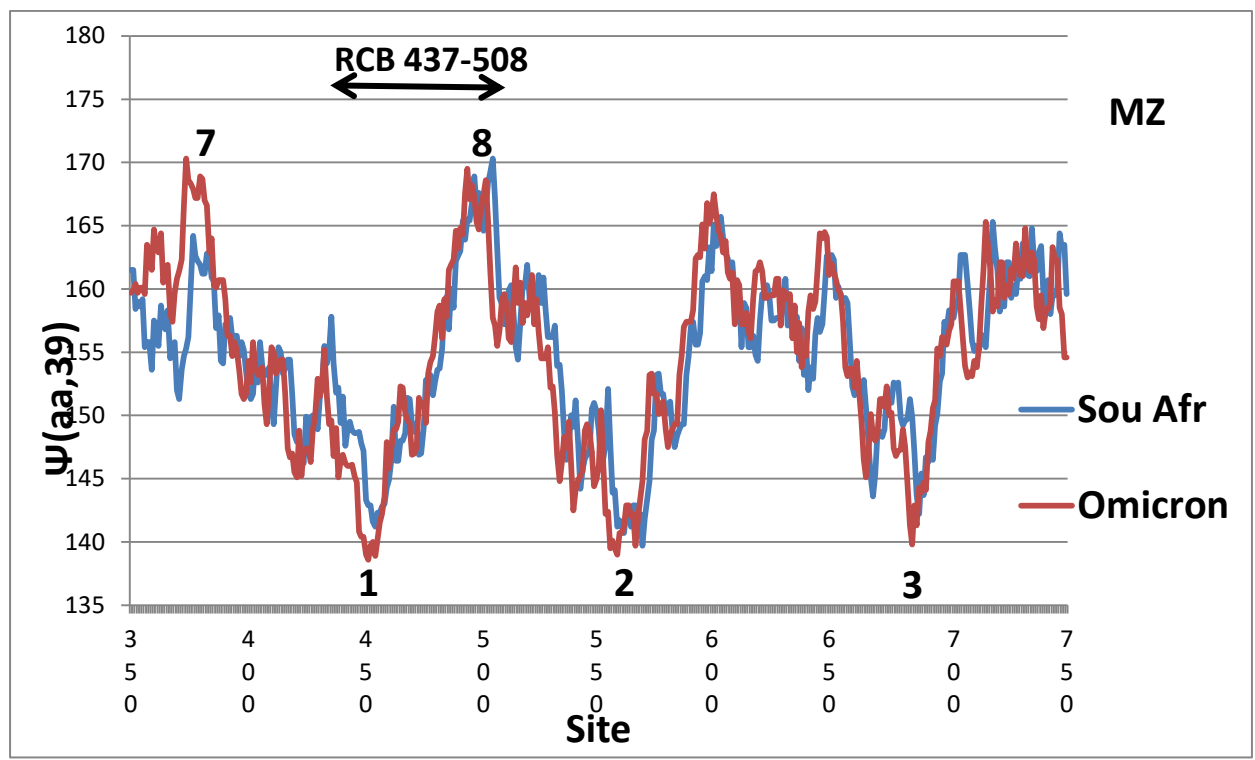

Fig. 2. Comparison of South Africa (2020) strain with Omicron strain (2021). The Ψ(aa,39) profiles are slightly misaligned for clarity. The hydrophilic edges 1-3 in South Africa have become more hydrophilic and are more level in Omicron. Peaks 7 and 8 were far from level in South Africa, but are 10x more level in Omicron.

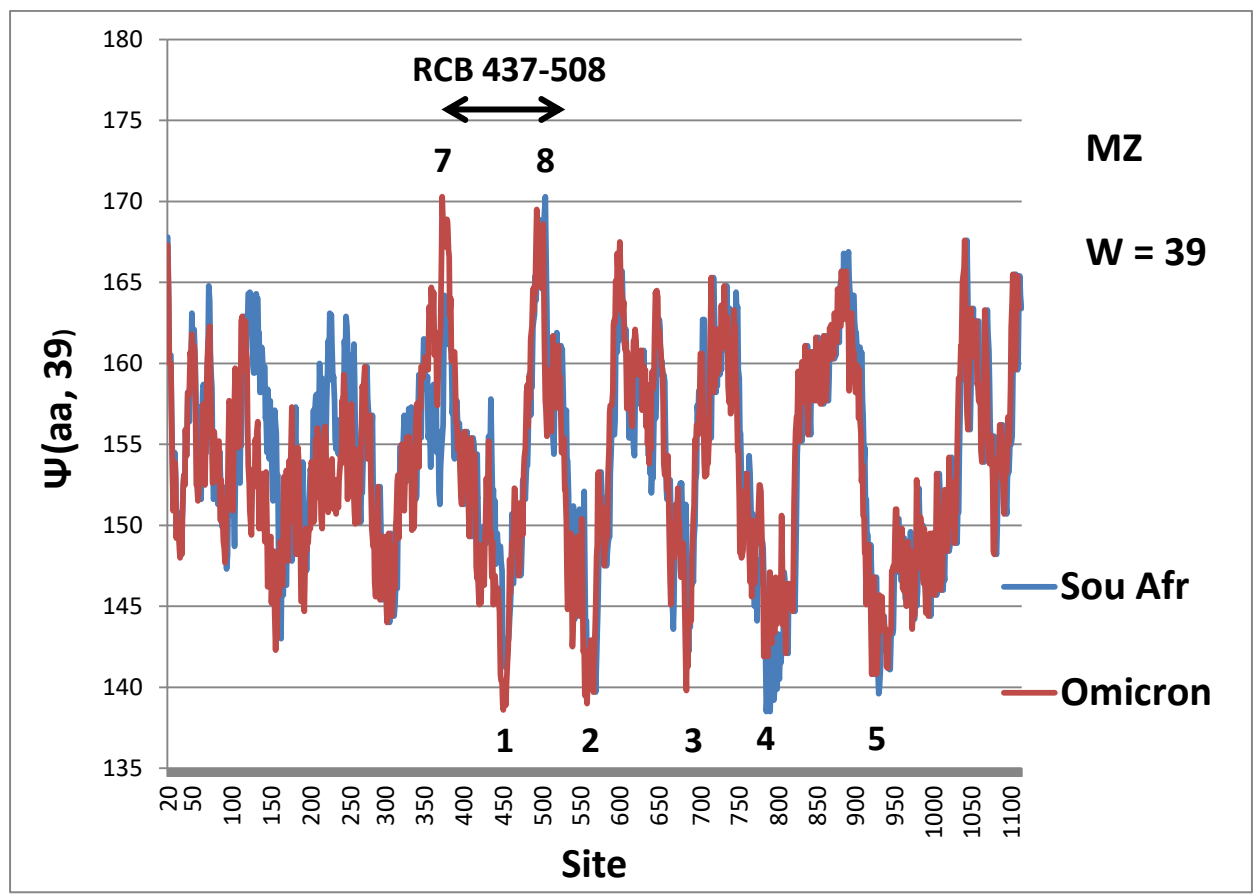

Fig. 3. Overall profiles of South Africa and Omicron. While the three hydrophilic edges 1-3, together with two hydrophobic edges, have become more level in Omicron, the additional S-2 hydrophilic edges 4,5 that were level in CoV-2 and Delta , are no longer level with 1-3 in Omicron. Thus the accelerated dynamics, which was previously spread over a wide range from sites 400 to 1200, is localized in the receptor region 350 - 600, and is correlated to the S1-S2 cleavage site 681-684 near edge 3. Note also that the now synchronized hydrophobic edges 7 (at 355) and 5 (at 477) are close to the narrowly defined receptor binding domain RCB [

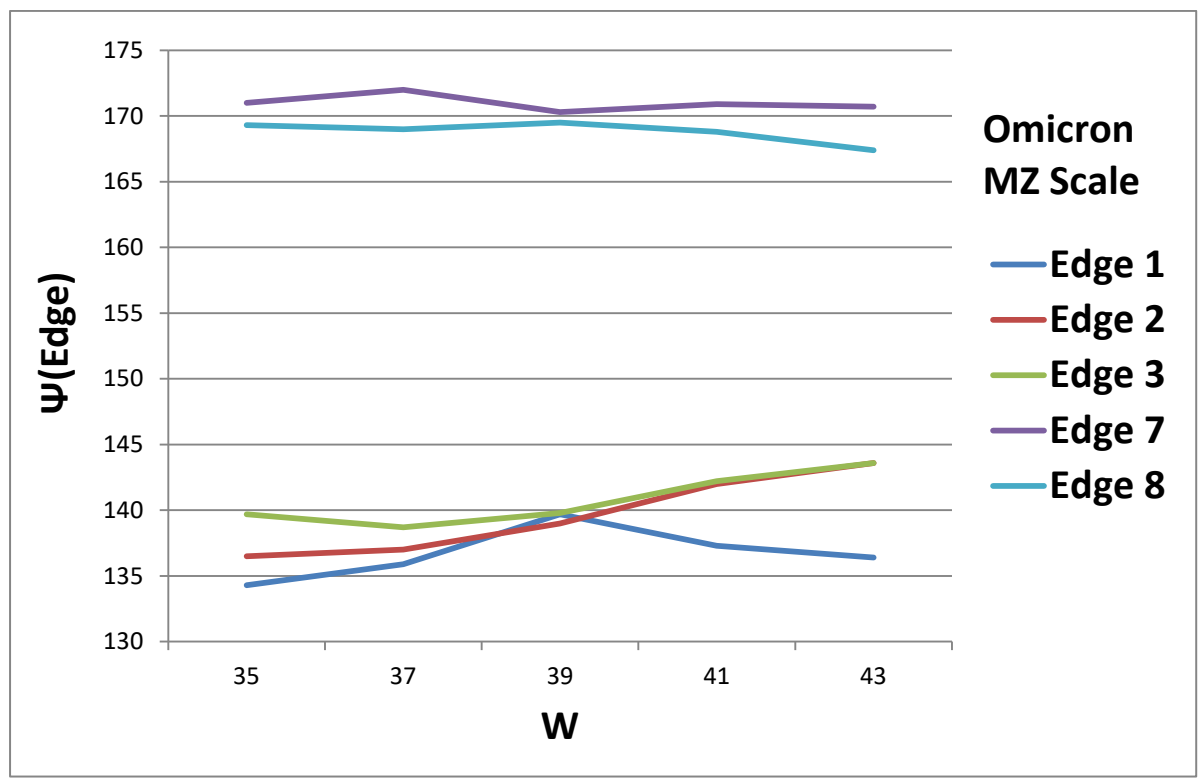

Fig. 4. Tuning of hydrophilic edges (1-3 in Figs. 2 and 3) and hydrophobic edges (7 and 8) with sliding window width W in Omicron. The leveling is best at W = 39 for both groups. The mean deviations 0.3 (philic) and 0.4 (phobic) at W = 39 are ~ 1% of the separation of the two groups.

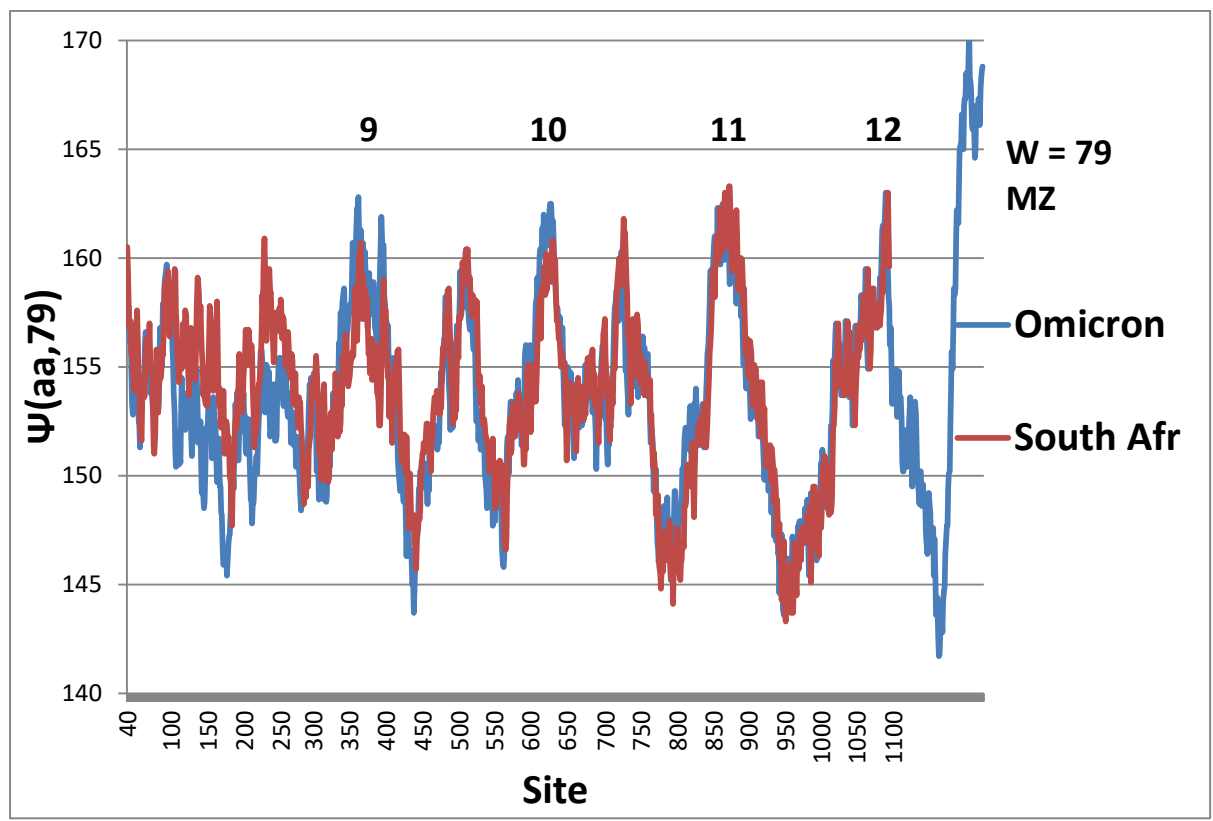

Fig. 5. Doubling W from 39 to 79 uncovers a new set of hydrophobic edges, 9-12, which are level in Omicron, but not in the South African strain.